\documentclass[runningheads]{llncs}
\usepackage[T1]{fontenc}
\usepackage{graphicx}
\makeatletter
\RequirePackage[bookmarks,unicode,colorlinks=true]{hyperref}%
\def\@citecolor{blue}%
\def\@urlcolor{blue}%
\def\@linkcolor{blue}%
\def\orcidID#1{\smash{\protect\raisebox{-1.25pt}{\protect\href{http://orcid.org/#1}{\includegraphics{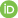}}}}}
\makeatother

\usepackage{morefloats} 

\usepackage{amsmath}
\newcommand{\emphTerminology}[1]{\textbf{#1}}

\newcommand{\M}{{\mathcal{M}}}

\newcommand{\A}{{\mathcal{A}}}
\newcommand{\B}{{\mathcal{B}}}

\newcommand{\partto}{\rightharpoonup}
\newcommand{\converges}{\ensuremath{\mathord{\downarrow}}}
\newcommand{\diverges}{\ensuremath{\mathord{\uparrow}}}

\newcommand{\alphabet}{\Sigma}

\newcommand{\languageof}[1]{\mathcal{L}(#1)}

\newcommand{\Hyp}{\mathcal{H}}
\newcommand{\Spec}{\mathcal{S}}

\newcommand{\MealyToDFAFiterau}[1]{\mathsf{MealyToDFA}(#1)}

\newcommand{\Complete}[1]{\mathsf{Complete}(#1)}
\newcommand{\Complement}[1]{(#1)^c}

\usepackage{amssymb}
\usepackage{bm}

\usepackage{nicefrac}

\usepackage[capitalize]{cleveref} 

\crefname{rule}{rule}{rules}
\crefname{IHItem}{IH item}{IH items}
\crefname{THMItem}{Theorem item}{Theorem items}
\crefname{LemmaItem}{Lemma item}{Lemma items}

\usepackage[x11names]{xcolor} 

\usepackage{csquotes}

\usepackage{marginnote} 

\usepackage{placeins} 

\usepackage{booktabs}
\usepackage[export]{adjustbox}
\usepackage{siunitx}
\usepackage{float}
\usepackage{caption}

\newcommand{\inputrotateddatatable}[3]{
  \begin{table}[H]
    \adjustbox{angle=90}{
      \begin{minipage}{\textwidth}
        \expandafter\label\expandafter{#2}%
        \input{#1}
        \parbox{15cm}{\centering \captionof{table}{#3}}%
      \end{minipage}
    }
  \end{table}%
}

\usepackage{tikz}
\usepackage{fancyvrb}
\usetikzlibrary{external, automata, positioning, arrows, decorations.markings}

\usetikzlibrary{intersections, shapes.geometric, calc,
	automata, arrows, decorations.pathreplacing,decorations.pathmorphing}
\usetikzlibrary{cd}
\usetikzlibrary{patterns, shapes} 
\usepackage{multirow}

\tikzset{
	initial text={},
	treenode/.style = {align=center, inner sep=0pt, text centered},
	basis/.style = {
		pattern=north east lines,
		pattern color=magenta!80!black!80!white,
	},
	frontier/.style={
		pattern=crosshatch dots,
		pattern color=yellow!80!black,
	},
	q0class/.style={
		pattern=vertical lines,
		pattern color=red!60!white,
	},
	q1class/.style={
		pattern=north west lines,
		pattern color=blue!60!white,
	},
	q2class/.style={
		pattern=crosshatch,
		pattern color=green!50!black!60!white,
	},
}

\tikzset{
    ->, 
    >=stealth, 
    node distance=1.5cm, 
    every state/.style={thick, minimum size=0.6cm, inner sep=0pt}, 
    initial text=, 
}
\usepackage{todonotes}

\begin{document}
\title{Systematic Evaluation of Black-Box Checking for Fast Bug Detection\thanks{This research was supported by NWO project OCENW.M.23.155 \enquote{Evidence-Driven Black-Box Checking (EVI)}.}}
%
%
\author{Bram Pellen\inst{1}\orcidID{0009-0009-3721-1964} \and
María Belén Rodr\'iguez\inst{2}\orcidID{0009-0009-7537-5222} \and 
Frits Vaandrager\inst{1}\orcidID{0000-0003-3955-1910} \and
Petra van den Bos\inst{2}\orcidID{0000-0002-9212-1525}
}
\authorrunning{Bram Pellen et al.}
%
\institute{Institute for Computing and Information Sciences, Radboud University, Nijmegen, the Netherlands \and
Formal Methods \& Tools, University of Twente, the Netherlands}
\maketitle              
\begin{abstract}
Combinations of active automata learning, model-based testing and model checking have been successfully used in numerous applications, e.g., for spotting bugs in implementations of major network
protocols and to support refactoring of embedded controllers.
However, in the large majority of these applications, model checking is only used at the very end, when no counterexample can be found anymore for the latest hypothesis model.
This contrasts with the original proposal of black-box checking (BBC) by Peled, Vardi \& Yannakakis, which applies model checking for \emph{all} hypotheses, also the intermediate ones.
In this article, we present the first systematic evaluation of the ability of BBC to find bugs quickly, based on 77 benchmarks models from real protocol implementations and controllers for which specifications of safety properties are available.
Our main finding are:
(a) In cases where the full model can be learned, BBC detects violations of the specifications with just 3.4\% of the queries needed by an approach in which model checking is only used for the full model. 
(b) Even when the full model cannot be learned, BBC is still able to detect many violations of the specification. In particular, BBC manages to detect 94\% of the safety properties violations in the challenging RERS 2019 industrial LTL benchmarks.
(c) Our results also confirm that BBC is way more effective than existing MBT algorithms in finding deep bugs in implementations.

\end{abstract}

\section{Introduction}
\label{sc:intro}

In their seminal paper from 1999 on \emph{black-box checking (BBC)}, Peled, Vardi \& Yannakakis \cite{PeledVY99,PeledVY02} combine three approaches for the verification of computer-based systems: 
\emph{model checking}, which checks if a known state diagram model conforms to a specification,
\emph{conformance testing} (a.k.a.\ \emph{model-based testing (MBT)}), which checks if a black-box system conforms with an abstract design, and \emph{active automata learning} (a.k.a.\ \emph{model learning}), which constructs a state diagram model of a black-box system by providing inputs and observing outputs. 
By combining these approaches, the authors obtain a method to check whether a black-box implementation satisfies a high-level specification.
Conceptually, BBC is just a form of model-based testing. However, by learning the state-transition behavior of the SUT during testing, BBC is often more effective than existing MBT algorithms in finding deep bugs in implementations for which only a high-level specification is available. 
We prefer the term black-box checking, but one could argue that the phrase \emph{learning-based testing} coined by Meinke \cite{MeinkeNS11,MeinkeS11} is more appropriate.

\begin{figure}[ht!]
\begin{center}
\tikzstyle{decision} = [diamond, draw, fill=blue!20, text width=3em, text badly centered, node distance=2cm, inner sep=0pt]
\tikzstyle{block} = [rectangle, draw, fill=blue!20, text width=4.1em, text centered, rounded corners, minimum height=2em]
\tikzstyle{line} = [draw, -latex']
\tikzstyle{io} = [trapezium, trapezium left angle=70, trapezium right angle=110, minimum width=1cm, minimum height=2em, text centered, draw=black, fill=lime!50]

\scalebox{1}{
\begin{tikzpicture}[node distance = 2cm, auto, every node/.style={}, every path/.style={}]
\node [block, minimum width=3cm, text width=3cm] (dispatcher) {Dispatcher};
\node [draw=none, above=0.5cm of dispatcher, node distance=8mm] (INIT) {};

\node [draw=none, below of=dispatcher] (center) {};
\node [block, minimum width=2cm, text width=2cm, left=0.6cm of center] (modelChecker) {Model Checker};
\node [block, minimum width=2cm, text width=2cm, left=1.5 cm of modelChecker] (learner) {Model Learner};
\node [block, minimum width=2cm, text width=2cm, right=0.6cm of center] (mbt) {Model-Based Tester};
\node [block, minimum width=1.5cm, text width=1.5cm, right=0.5 cm of mbt] (monitor) {Runtime Monitor};

\node [block, below of=center, minimum width=2cm, yshift=-1cm,text width=2.75cm] (obtree) {Adapter};
\node [block, below=0.75cm of obtree, minimum width=2cm, text width=2cm, color=black] (sut) {\textcolor{white}{SUT}};

\path [line] ([xshift=-0.25cm]INIT.south)-- node[left] {$\cal S$} ([xshift=-0.25cm] dispatcher.north);
\path [line] ([xshift=0.25cm]dispatcher.north)-- node[right] {(no) bug found} ([xshift=0.25cm] INIT.south);

\path [line] (dispatcher.west) -- node[above] {go} (learner.north);
\path [line] ([xshift=-0.25cm]dispatcher.south) -- node[below] {$\cal S$} ([xshift=0.25cm]modelChecker.north);

\path [line] ([xshift=0.7cm]monitor.north) -- node[right, align=center,xshift=1cm,yshift=-0.25cm] {bug\\found} ([yshift=0.25cm]dispatcher.east);
\path [line] ([yshift=-0.25cm]dispatcher.east) -- node[below] {$\Spec$} ([xshift=-0.2cm]monitor.north);

\path [line] ([yshift=0.15cm]learner.east) -- node[above] {$\Hyp$} ([yshift=0.15cm]modelChecker.west);
\path [line] ([yshift=-0.15cm]modelChecker.west) -- node[below] {CE for~$\Hyp$} ([yshift=-0.15cm]learner.east);

\path [line] ([xshift=-0.75cm]learner.south) -- node[left] {$\sigma_\mathit{in}$} ([yshift=-0.2cm, xshift=-0.05cm]obtree.west);
\path [line] ([yshift=0.1cm, xshift=0.05cm]obtree.west) -- node[right] {$\sigma_\mathit{out}$} ([xshift=-0.25cm]learner.south);

\path [line] ([xshift=-0.25cm]modelChecker.south) -- node[left] {$\sigma_\mathit{in}$} ([xshift=-1cm]obtree.north);
\path [line] ([xshift=-0.5cm]obtree.north) -- node[right] {$\sigma_\mathit{out}$} ([xshift=0.25cm]modelChecker.south);

\path [line] ([xshift=-0.25cm]mbt.south) -- node[left] {$\sigma_\mathit{in}$} ([xshift=0.25cm]obtree.north);
\path [line] ([xshift=0.75cm]obtree.north) -- node[right] {$\sigma_\mathit{out}$} ([xshift=0.25cm]mbt.south);

\path [line] (modelChecker) -- node[below] {$\Hyp$} node[above] {$[\Hyp\models\Spec]$} (mbt);

\path [line] ([xshift=-0.75cm]mbt.south) -- node[above, xshift=-4.5cm, yshift=-0.35cm]{CE for~$\Hyp$} ++(-0.4,-0.6) -- ([xshift=1.15cm, yshift=-0.6cm]learner.south) -- ([xshift=0.35cm]learner.south);

\path [line] ([xshift=-0.75cm]mbt.north) -- node[right,align=center] {no bug\\found} ([xshift=0.25cm]dispatcher.south);

\path [line] (obtree.east) -- node[below, xshift=0.6cm] {$\sigma_\mathit{in}/ \sigma_{out}$} (monitor.south);

\path [line] ([xshift=-0.25cm]obtree.south) --  ([xshift=-0.25cm]sut.north);
\path [line] ([xshift=0.25cm]sut.north) -- ([xshift=0.25cm]obtree.south);

\end{tikzpicture}
}
\end{center}
\caption{Black-box checking for specification $\mathcal{S}$ and system under test SUT.}
\label{Fig:BBC}
\end{figure}
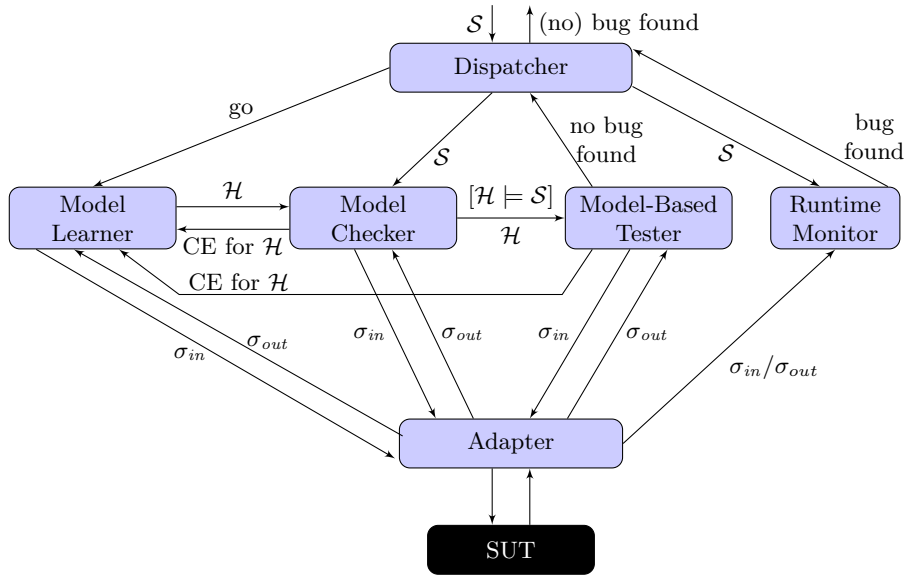

Figure~\ref{Fig:BBC} schematically shows how BBC works.
First of all, it assumes an implementation (of a black-box system), referred to as the \emph{System Under Test (SUT)}, that needs to be analyzed.
It also assumes a (partial) specification $\Spec$, usually a conjunction of a number of requirements on the behavior of the SUT.
An \emph{Adapter} translates the abstract input and output symbols from the specification to concrete inputs and outputs that are accepted/produced by the SUT. The adapter makes BBC scalable to realistic applications, by using powerful abstractions \cite{AJUV15,VW23}.
A \emph{Dispatcher} orchestrates the activities of the different analysis tools used in the BBC approach.
The dispatcher starts the analysis by activating the \emph{Model Learner} \cite{Angluin87,Vaandrager17,HowarS18}, which builds a state-transition model $\Hyp$ (an \emph{hypothesis}) from results of \emph{learning queries} (a.k.a.\ \emph{output queries}): the learner sends sequences $\sigma_\mathit{in}$ of inputs to the adapter/SUT, and in return receives sequences $\sigma_\mathit{out}$ of outputs. The hypothesis $\Hyp$ aims to describe how the SUT actually works. 
Whereas specification model $\Spec$ will often be small (e.g., a two state model describing that a \emph{response} may only occur after a \emph{request}), the hypothesis $\Hyp$ will typically be much bigger, because the learning algorithm discovers --- through systematic exploration --- that states of the SUT reached via certain input sequences have different input-output behavior.
Subsequently, BBC runs a \emph{Model Checker} \cite{BK08,HandbookModelChecking} to check if hypothesis $\Hyp$ \emph{satisfies} specification $\Spec$, writen $\Hyp \models \Spec$, meaning that all behaviors of $\Hyp$ are allowed by $\Spec$.
Two cases are considered:
\begin{enumerate}
	\item 
	If $\Hyp \models \Spec$ then a \emph{Model-Based Tester} \cite{LeeY96,Tret08} will run \emph{testing queries} to check whether $\Hyp$ also is a correct model of the SUT: it sends sequences $\sigma_\mathit{in}$ of inputs to the adapter/SUT, and checks if the resulting sequence $\sigma_\mathit{out}$ of outputs agree with the prediction of $\Hyp$.  If a test fails, then the model learner uses this test to improve $\Hyp$.  If all tests pass then BBC terminates and reports that no evidence was found that the SUT violates specification $\Spec$.
	\item 
	If $\Hyp\not\models\Spec$ then there is a counterexample $\sigma_\mathit{in}$ for which $\Hyp$ produces an output not allowed by $\Spec$.
	BBC then performs an additional learning query $\sigma_\mathit{in}$ on the adapter/SUT.
	If the resulting output $\sigma_\mathit{out}$ agrees with the prediction by $\Hyp$ then the SUT violates $\Spec$.
	Otherwise, $\Hyp$ is not a correct model of the SUT, and the model learner uses the counterexample to further improve $\Hyp$.
\end{enumerate}
Finally, we  included a \emph{Runtime Monitor} \cite{LeuckerS09} in Figure~\ref{Fig:BBC}, as a minor extension of the BBC framework presented in \cite{PeledVY99,PeledVY02}.\footnote{Monitors have also been added to BBC by Meijer \& Van de Pol \cite{MeijerP19} but with a different purpose.  They study BBC for general LTL formulas and propose to let the model checker consider the safety portion of an LTL property first and derive simpler counterexamples using monitors.  In the benchmarks that we consider in this paper all requirements are safety properties.}
The monitor checks, for all learning queries performed on the SUT, whether the output $\sigma_\mathit{out}$ produced in response to an input $\sigma_\mathit{in}$ is allowed by specification $\Spec$.  If an output violates $\Spec$ then a bug is reported.  Use of a runtime monitor allows us to detect bugs faster: when a bug is encountered, it is reported immediately, rather than using it to further improve hypothesis model $\Hyp$.

Combinations of model learning, model-based testing and model checking have been successfully used in numerous applications, e.g.,
for spotting bugs in implementations of major network
protocols~\cite{RuiterP15,FiterauBrosteanJV16,DBLP:conf/spin/Fiterau-Brostean17,FiterauBrosteanH17,FiterauBrosteanJMRSS20,FerreiraBDS21} and to support refactoring of embedded controllers \cite{SmeenkMVJ15,SHV16,YangASLHCS19}.
We refer to~\cite{Vaandrager17,HowarS18,AichernigMMTT18} for surveys and further references.
However, in the large majority of these applications, model checking is only used at the very end, when no counterexample can be found anymore for the latest hypothesis model (exceptions are e.g.\ \cite{MeinkeS11,MeinkeNS11,MeijerP19}). This contrasts with the original BBC proposal of \cite{PeledVY99,PeledVY02} which applies model checking for \emph{all} hypotheses, also the intermediate ones.
This is remarkable since, as observed by \cite{PeledVY99,PeledVY02}, ``intuitively it is clear that in many cases this method [only checking the final hypothesis] can be wasteful in that it does not take advantage of the property to avoid doing a complete identification''.
In a black-box setting we can never be sure about the correctness of learned models unless we are willing to make strong (typically unrealistic) assumptions, such as a bound on the number of states of the system. Therefore, rather than learning models, the main goal of BBC is to find bugs in implementations as fast as possible.

Some serious studies have been carried out to benchmark combinations of learning and testing algorithms for BBC \cite{GarhewalD23,AichernigTW24}, but (surprisingly) there is no systematic evaluation of the ability of BBC to find bugs quickly.
The objective of this article is to provide such an evaluation.
In particular, we evaluate:
\begin{enumerate}
\item[a)] 
The effectiveness of model checking all hypotheses in BBC.
Can we confirm the conclusion of Meinke \& Sindhu \cite{MeinkeS11}, obtained for a single benchmark of an elevator system, that the time required by BBC to discover a bug in the SUT is between 0.003\% and 7\% of the total time needed to completely learn the full model? 
\item[b)]
The effectiveness of BBC to detect specification violations in situations where the full model cannot be learned.
\item[c)]
The ability of BBC to find bugs when compared with traditional model-based testing algorithms such as described in \cite{LeeY96,Tret08,DorofeevaEMCY10,BosV21}.
\end{enumerate}

In our experiments we simulate the SUT from benchmark models that were obtained using automata learning in previous work. Such models are available on the Automata Wiki \cite{NeiderSVK97}\footnote{\url{https://automata.cs.ru.nl/}}.
From this Wiki, we selected 77 benchmarks models with the following characteristics:
\begin{itemize}
	\item 
	All models were obtained via BBC from real protocol implementations and embedded controllers. 
	\item 
	All models are deterministic Mealy machines in which each input from a finite  set $I$ triggers a sequence of outputs taken from a finite set $O$.
	\item 
	For all models, a specification was available consisting of a number of safety properties specified either as a DFA over alphabet $I \cup O$, or as an LTL formula that we converted to a  DFA over alphabet $I \times O^*$ using Spot \cite{duret.22.cav}.
	Even though the specifications that we consider are DFAs, multiple outputs may be enabled in a single state, meaning that they allow for \emph{observable nondeterminism} in the sense of \cite{PY06,BosJM19}.
\end{itemize}
The rest of this article is structured as follows.
Section 2 recalls the basic notations and definitions related to DFAs and Mealy machines that we will use.
Section 3 briefly introduces the case studies from which we obtained our benchmarks models and specifications.
Section 4 discusses our experimental setup, both for black-box checking and for model-based testing.
The results from our experiments are presented in Section 5.
Finally, Section 6 presents our conclusions and discusses directions for future research.
All data and software required to reproduce our experiments will be made available together with the final version of this article.

\section{Preliminaries}

In this preliminary section, we first fix notation for partial maps and sequences, and then for DFAs and Mealy machines. Next, we describe how DFAs can be used to specify properties of Mealy machines, and a simple model checking approach for this setting. Finally, we briefly discuss our MBT approach.

\subsection{Partial maps and sequences}
We write $f \colon X \partto Y$ to denote that $f$ is a partial function from $X$
to $Y$ and write $f(x) \converges$ to mean that $f$ is defined on $x$, that is,
$\exists y \in Y \colon f(x)=y$, and conversely write $f(x)\diverges$ if $f$ is
undefined for $x$.
Function $f \colon X \partto Y$ is \emph{total} if $f(x) \converges$, for all $x \in X$.
Often, we identify a partial function $f \colon X \partto Y$ with the set $\{ (x,y) \in X \times Y \mid f(x)=y \}$. The composition of partial
maps $f\colon X\partto Y$ and $g\colon Y\partto Z$ is denoted by $g\circ
f\colon X\partto Z$, and we have $(g\circ f)(x)\converges$ iff $f(x)\converges$
and $g(f(x)) \converges$.
We use the Kleene equality on partial functions, which states that on a given argument either both functions are undefined, or both are defined and their values on that argument are equal. 


An \emphTerminology{alphabet} $\Sigma$ is a finite set of symbols.
A \emphTerminology{word} over alphabet~$\alphabet$ is a finite sequence of symbols from $\Sigma$.
We write $\epsilon$ to denote the empty word, and $a$ to denote the word consisting of a symbol $a \in \Sigma$.
If $w$ and $w'$ are words over $\Sigma$ then we write $w \; w'$ to denote their \emphTerminology{concatenation}.
We write $\Sigma^*$ to denote the set of all words over $\Sigma$, and $\Sigma^+$ to denote set $\Sigma^* \setminus \{ \epsilon \}$.
A word $v$ is a \emphTerminology{prefix} of a word $w$ if there exists a word $u$ such that $w = v \; u$. Similarly,
$v$ is a \emphTerminology{suffix} of  $w$ if there exists a word $u$ such that $w = u \; v$.
A \emphTerminology{language} over $\Sigma$ is a subset of $\Sigma^*$.
If $L$ is a language over $\Sigma$, then we define $\Complement{L}$ as the language $\Sigma^* \setminus L$.
A language $L$ is \emphTerminology{prefix closed} if  $w \in L$ implies  $v \in L$, for any prefix $v$ of $w$.

\subsection{DFAs and Mealy machines}

\begin{definition}[DFA]
    A (partial) \emphTerminology{deterministic finite automaton (DFA)} is a five-tuple~$\A = (Q, \Sigma, \delta, q^0, F)$, where~$Q$ is a finite set of \emphTerminology{states}, $\Sigma$ is a set of \emphTerminology{input symbols}, $q^0 \in Q$ is the \emphTerminology{initial} state, $F \subseteq Q$ is a set of \emphTerminology{final} states, and $\delta : Q \times \Sigma \partto Q$ is a (partial)  \emphTerminology{transition function}.
    The transition function is inductively extended to words over $\Sigma$ in the standard way, (for $w \in \Sigma^*$ and $a \in \Sigma$):
    \begin{eqnarray*}
    	\delta(q, \epsilon) & = & q\\
    	\delta(q, w \; a) & = & \delta(\delta(q, w), a)
    \end{eqnarray*}
    A word $w \in \Sigma^*$ is \emphTerminology{accepted} by $\A$ if $\delta(q^0, w) \in F$, and \emphTerminology{rejected} by $\A$ if $\delta(q^0, w) \not\in F$.
    The \emphTerminology{language} accepted by $\A$, denoted $\languageof{\A}$, is the set of all words accepted by $\A$.
    Two DFAs $\A$ and $\A'$ are \emphTerminology{equivalent}, denoted $\A \approx \A'$, if they accept the same language.
    A DFA $\A$ is \emphTerminology{prefix closed} if, for each state $q$ and input $a$,  $\delta(q,a) \in F$ implies $q \in F$.
    DFA $\A$ is \emphTerminology{complete} if transition function $\delta$ is total.
\end{definition}

It is easy to see that
if $\A$ is prefix closed then $\languageof{\A}$ is prefix closed.
Each DFA can be turned into a complete DFA by adding a fresh (nonfinal) \emphTerminology{sink state} and adding, for each missing transition, a transition to that sink state.

\begin{definition}[Completion of a DFA]
Let $\A = (Q, \Sigma, \delta, q^0, F)$ be a DFA.
Then $\Complete{\A}$ is the complete DFA $(Q', \Sigma, \delta', q^0, F)$, where
$Q'  =  Q \cup \{ q_s \}$ and, for all $q \in Q'$ and $a \in \Sigma$,
\begin{eqnarray*}
	\delta'(q, a) & = & \begin{cases}
		\delta(q,a) & \mbox{if } \delta(q,a) \converges\\
		q_s & \mbox{otherwise}
	\end{cases}
\end{eqnarray*}
\end{definition}

It is straightforward to verify that $\A \approx \Complete{\A}$.

Whereas the output of a DFA is limited to a binary signal (accept/reject), a Mealy machine associates a more general output to each transition.  The definition below, adapted from \cite{DBLP:conf/ndss/Fiterau-Brostean23}, associates a sequence of outputs from some alphabet to each transition. Such machines are quite useful to model the behavior of protocol entities.

\begin{definition}[Mealy machine]
    A (partial) \emphTerminology{Mealy machine} is a tuple $\M = (I, O, Q, q_0, \delta, \lambda)$, where~$I$ and $O$ are alphabets of \emphTerminology{input} and \emphTerminology{output symbols}, respectively, $Q$ is a set of \emphTerminology{states} containing the \emphTerminology{initial} state $q_0$, $\delta : Q \times I \partto Q$ is a (partial) \emphTerminology{transition function}, and $\lambda : Q \times I \partto O^*$
    is a (partial) \emphTerminology{output function}. We require, for all $q \in Q$ and $i \in I$, that $\delta(q,i)\converges$  iff $\lambda(q,i)\converges$. A Mealy machine is \emphTerminology{complete} if $\delta$ (and hence $\lambda$) is total.
\end{definition}

\subsection{Model checking}

As part of BBC, the model checker verifies if a hypothesis $\Hyp$ satisfies specification $\Spec$. Our hypotheses are Mealy machines, while we use Deterministic Finite Automata (DFA) for specifications. Inspired by \cite{DBLP:conf/ndss/Fiterau-Brostean23}, we translate our Mealy machines into DFAs, such that satisfaction of $\Hyp$ to $\Spec$ can be checked using standards available for DFAs \cite{HU79}.

To translate from Mealy machines to DFAs, the idea is to insert auxiliary states $(w, q)$ in between the source $q'$ and target $q$ of a transition, in which first the outputs from $w$ are performed before jumping to state $q$.

\begin{definition}[From Mealy machines to DFAs]
	Let $\M = (I, O, Q, q_0, \delta, \lambda)$ be a Mealy machine, with $I \cap O = \emptyset$.
	Then $\MealyToDFAFiterau{\M}$ is the partial DFA $(Q', I \cup O, \delta', q^0, Q')$ where, for all $q \in Q$, $i \in I$, $o, o'\in O$ and $w \in O^*$: 
	\begin{eqnarray*}
		Q'& = & Q \cup Q_{\mathit{aux}}\\
		Q_{\mathit{aux}} & = & \{ (v, q) \in O^+ \!\!\times Q \mid \exists q' \in Q , j \in I : \delta(q', j) = q  \mbox{ and } v \mbox{ suffix of } \lambda(q', j)   \}\\
		\delta'(q,i) & = & \begin{cases}
			\delta(q,i) & \mbox{if } \lambda(q,i) = \epsilon\\
			(\lambda(q,i), \delta(q, i)) & \lambda(q,i) \in O^+\\
			\mbox{undefined} & \lambda(q,i)\diverges
			\end{cases}\\
		\delta'(q,o)\diverges\\
		\delta'((o \; w, q), o') & = & \begin{cases}
			\mbox{undefined} & \mbox{if } o \neq o'\\
			q & \mbox{if } o=o' \mbox{ and } w = \epsilon\\
			(w, q) & \mbox{otherwise}
		\end{cases}\\
		\delta'((o \; w, q), i)\diverges
	\end{eqnarray*}
\end{definition}

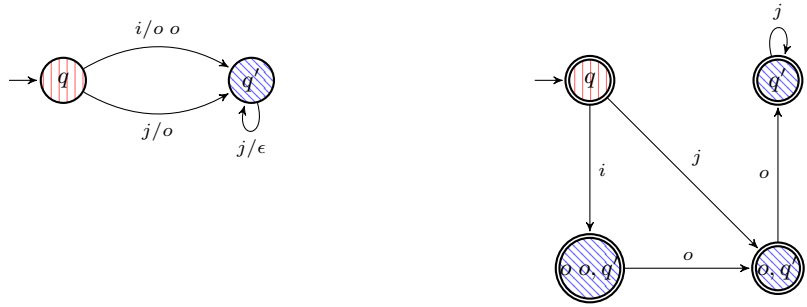
\begin{figure}[h]
	\begin{center}
		\begin{tikzpicture}[->,>=stealth',shorten >=1pt,auto,node distance=2.5cm,main node/.style={circle,draw,font=\sffamily\large\bfseries},
			]
			\def\yoffset{8mm}
			\node[initial,state,q0class] (0) {{$q$}};
			\node[state,q1class] (1) [right of=0] {{$q'$}};

			\node[initial,state,q0class,accepting] [right of=1,xshift=2cm](2) {{$q$}};
			\node[state,q1class,accepting] (3) [below of=2] {{$o \; o, q'$}};
			\node[state,q1class,accepting] (4) [right of=3] {{$o, q'$}};
			\node[state,q1class,accepting] (5) [above of=4] {{$q'$}};

			\path[every node/.style={font=\sffamily\scriptsize}]
			(0) edge[bend left] node[sloped,above] {$i/o \; o$} (1)
				edge[bend right] node[sloped,below] {$j/o$} (1)
			(1) edge[loop below] node {$j/ \epsilon$} (1)
			(2) edge node {$i$} (3)
				edge node {$j$} (4)
			(3) edge node {$o$} (4)
			(4) edge node {$o$} (5)
			(5) edge [loop above] node {$j$} (5);
		\end{tikzpicture}
		\caption{A Mealy machine (left) and the corresponding DFA (right).}
		\label{Fig:MealyToDFA}
	\end{center}
\end{figure}

Figure~\ref{Fig:MealyToDFA} gives an example of our translation from Mealy machines to DFAs.
Note that $\MealyToDFAFiterau{\M}$ is a labelled transition system with inputs and outputs in the sense of Tretmans~\cite{Tret08}, but of a restricted form.  States of the DFA either have a \emph{single} outgoing output transition, or zero or more outgoing input transitions.
Moreover, at most finitely many consecutive output transitions are possible from any state.
It is easy to see that, given $I$ and $O$, we may fully retrieve $\M$ from $\MealyToDFAFiterau{\M}$.
Our translation improves on the one presented in \cite{DBLP:conf/ndss/Fiterau-Brostean23}, which turns the Mealy machine of Figure~\ref{Fig:MealyToDFA} into a DFA with 5 states.

\begin{definition}[Specification]\label{def:specification}
	\label{def specification}
	Consider a Mealy machine $\M$ with inputs $I$ and outputs $O$.
	A \emphTerminology{specification} for $\M$ is a prefix closed DFA $\Spec$ over alphabet $I \cup O$.
	We say that $\M$ \emph{satisfies} specification $\Spec$ if
	 $\languageof{\MealyToDFAFiterau{\M}} \subseteq \languageof{\Spec}$.
\end{definition}

The specifications for our benchmark models need to be massaged a bit to turn them into the above format.
For the BLE and RERS case studies the specifications are LTL formulas. Using the Spot tool \cite{duret.22.cav}, we translate these to equivalent DFAs over alphabet $I \times O$.
By splitting each $(i, o)$-transition into an $i$-transition followed by an $o$-transition, we can translate these DFA into (prefix-closed) DFAs over alphabet $I \cup O$.
For the SSH and DTLS case studies, the specifications are given in terms of ```bug automata''. These are DFAs over alphabet $I \cup O$, but with the roles of final/nonfinal states interchanged: accepting runs correspond to undesired behavior of the SUT.  By complementing these bug automata, we obtain specifications in the sense of our Definition~\ref{def specification}.

A model checker can efficiently check that $\M$ satisfies $\Spec$, using the fact that for all DFAs $\A$ and $\B$, $\languageof{\A} \subseteq \languageof{\B}$ iff
$\languageof{\A} \cap \Complement{\languageof{\B}} = \emptyset$.
Both intersection and complement can be easily defined for DFAs:

\begin{definition}[Complement]
	Let $\A = (Q, \Sigma, \delta, q^0, F)$ be a complete DFA.  Then the \emphTerminology{complement} of $\A$, denoted $\Complement{\A}$, is the DFA $\A = (Q, \Sigma, \delta, q^0, Q \setminus F)$.
\end{definition}

\begin{definition}[Product]
	Let $\A_1 = (Q_1, \Sigma, \delta_1, q^0_1, F_1)$ and $\A_2 = (Q_2, \Sigma, \delta_2, q^0_2, F_2)$ be two DFAs.
	Then the \emphTerminology{product} of $\A_1$ and $\A_2$, denoted $\A_1 \parallel \A_2$, is the DFA ~$(Q_1 \times Q_2, \Sigma, \delta, (q^0_1, q^0_2), F_1 \times F_2)$, where for all $q_1 \in Q_1$, $q_2 \in Q_2$ and $a \in \Sigma$,
	\begin{eqnarray*}
		\delta((q_1, q_2), a) & = & (\delta_1(q_1, a), \delta_2(q_2, a)).
	\end{eqnarray*}
\end{definition}

It is well-known that for complete $\A$,
$\languageof{\Complement{\A}} = \Complement{\languageof{\A}}$
and, for DFAs $\A_1$ and $\A_2$ with the same alphabet,
$\languageof{\A_1 \parallel \A_2} = \languageof{\A_1} \cap \languageof{\A_2}$, see e.g. \cite{HU79}.
Thus $\M$ satisfies $\Spec$ iff
$\languageof{\MealyToDFAFiterau{\M} \parallel \Complement{\Complete{\Spec}}} = \emptyset$.
Emptiness of the language accepted by a DFA can be decided in time linear in the size of the DFA \cite{HU79}.

\subsection{Model-Based Testing}

In this paper, Model-Based Testing (MBT) is used in two different ways. First, MBT is used as part of the BBC procedure, namely to check conformance of the current hypothesis to the SUT. Secondly, we use MBT as a standalone method to find bugs in the SUT. This way, we can  compare the bug finding capabilities of BBC with standalone MBT.
In MBT, tests are derived from a model. In MBT as part of BBC this model is the hypothesis. Standalone MBT uses the specification as its model: it uses the prefix-closed DFAs over the alphabet $I \cup O$, where a transition is either labeled with an input or an output label.
Therefore these DFAs can be trivially translated  to a Labeled Transition Systems, for which test derivation is defined as usual \cite{Tret08}, i.e. a test either:
\begin{itemize}
 \item  supplies an input specified by the model
 \item waits to observe an output -- if the received output is not the output specified by the model, the test stops with verdict fail
 \item stops with verdict pass, when the stop condition, e.g. number of test steps, has been satisfied.
\end{itemize}
The test keeps track of the current state of the model according to the supplied inputs and observed outputs.
For selecting the inputs of a test, existing MBT test generation strategies can be used, e.g. selecting an input randomly, or selecting a different input than already tried. 



\section{Case Studies}\label{sec:case-studies}
We selected case studies from several published papers for which both models of SUTs were available, as well as properties that could be converted to DFAs.
The SUTs were available on the automata wiki; corresponding specifications were obtained from the paper and corresponding artefacts.
The case studies vary in size and application area, so that we have a good basis to draw conclusions.

\paragraph{BLE.}
In~\cite{DBLP:conf/nfm/PferscherA22}, Pferscher et al. use an active automata learning approach to look for anomalies and security vulnerabilities in five distinct System on Chip (SoC) Bluetooth Low Energy (BLE) devices, all of which implement the BLE 5 standard. Pferscher et al. found several crashes, anomalies, and a security vulnerability with their fuzzing-enhanced learning process.

\paragraph{SSH (LTL).}
In~\cite{DBLP:conf/spin/Fiterau-Brostean17}, Fiterau{-}Brostean et al. use active automata learning and model checking to verify several security and functional requirements against three SSH server implementations (Bitvise 7.23, Dropbear v2014.65 and OpenSSH 6.9p1-2). They use the NuSMV~\cite{DBLP:conf/cav/CimattiCGGPRST02} model checker to automatically check their LTL formalizations of 12 requirements imposed by RFCs that describe the second version of the SSH protocol (i.e., SSHv2) against the final Mealy machine models they learned for their selected implementations. They found that each of the implementations under test violated at least one of their selected functional requirements.

\paragraph{SSH (DFA).}
In~\cite{DBLP:conf/ndss/Fiterau-Brostean23}, Fiterau-Brostean et al., describe the bugs and vulnerabilities they look for as deterministic finite automata (DFAs), that they check during automata learning. They apply this on three more recent SSH server implementations
(BitVise-8.49, DropBear-v2020.81 and OpenSSH-8.8p1).

\paragraph{DTLS (Client \& Server).}
Fiterau-Brostean et al., also apply their method from \cite{DBLP:conf/ndss/Fiterau-Brostean23} to 16 DTLS client implementations and 17 DTLS server implementations (there were multiple versions of the same DTLS implementations).

\paragraph{RERS.}
The Rigorous Examination of Reactive Systems (RERS)-challenges are semi-annual open competitions in which participants attempt to solve a given set of reachability and verification problems.
The goal behind these competitions is to encourage the use of new combinations of various tools and approaches and to provide a basis of comparison for the approaches used by the participants. The 2019-edition of the RERS challenge~\cite{DBLP:conf/tacas/JasperMMSHSSHSK19} contained an industrial track with benchmark programs that are based on Mealy machine models of controller software provided by the company ASML.
We used all thirty SUTs of the RERS 2019 industrial LTL challenge, along with their accompanying LTL specifications.\footnote{\url{https://rers-challenge.org/2019/index.php?page=industrialProblemsLTL\#}}

\section{Experimental Setup} 

\subsection{Black-box Checking}

We implemented our black-box checking approach in Python 3.11.12 on top of version 1.5.1 of the AALpy~\cite{aalpy} library for active automata learning. We used version 2.13 of the LTL and~$\omega$-automata manipulation and model checking library Spot~\cite{duret.22.cav} to convert the LTL properties into monitors.
Additionally, for performing experiments, we used Docker\footnote{\url{https://www.docker.com/}}.

We used the active automata learning algorithm~$L^\#$ because it is competitive with that of several alternatives~\cite{lsharp}. 
We use separating sequences for~$L^\#$'s extension rule because this is the default in AALpy. For most experiments, we use Adaptive Distinguishing Sequences (ADS) for the separation rule as this is AALpy's default setting. The exception is RERS, where we used separating sequences because we found that using ADS with experiments that were otherwise unchanged made the hypothesis refinements and thereby the BBC process in the RERS models considerably slower.

We fixed the model-based tester for BBC to Hybrid-ADS \cite{SmeenkMVJ15} because it is a recent, state-of-the-art approach that is available as a tool~\footnote{\url{https://gitlab.science.ru.nl/bharat/hybrid-ads}}. 
We based our configuration for Hybrid-ADS on~\cite{lsharp}. As such, we set the operation mode to~\enquote{random}, the prefix mode to \enquote{buggy}, the number of extra states to check for (minus 1) to 10, and the expected number of random infix symbols to 10. We repeat each of our experiments for 50 random seeds to account for the randomness introduced by our use of Hybrid-ADS. We measure the time that each seed takes with the \texttt{perf\_counter\_ns}-function from the \texttt{time}-module of Python's standard library.

Hybrid-ADS produces an unbounded number of testing queries when used in random mode. Performing another correctness check when the current hypothesis is equivalent to the SUT would take an infinite amount of time. In~\cite{lsharp}, the authors skip this final correctness check if they determine that the current hypothesis is already equivalent to the SUT. We follow this approach for most of our case studies, since the number of testing queries that one would perform to look for a difference between a correct hypothesis and the SUT is always both finite, and ultimately arbitrary. The exception is the RERS case study, for which we perform at most 1e6 testing queries per correctness check. We did so to reduce the time that we need to run the RERS experiments, several of which would take at least an hour for a single seed when we used unlimited testing queries. Budgeting the number of queries brings with it the risk of not learning the complete models, which could lead to bugs remaining undetected. We selected the maximum of 1e6 testing queries as a trade-off between the number of properties for which we found counterexamples and the time that we would spend on each experiment. We determined the number itself based on experimentation with a subset of the RERS benchmarks.



We followed~\cite{KrugerJR24}'s example in using a step budget to further reduce the time taken by certain BBC experiments. If it is about to exceed this budget, BBC first terminates its current operation, and then uses the model checker to look for counterexamples for the properties for which it hadn't found one. This use of the model checker does imply that the step budget can be exceeded, since the model checker will still verify any counterexamples that it finds against the SUT. Whenever we used step budgets for an SUT, we did so because we found that it should take us considerably more than 48 hours to obtain all of our results for it. We then used a range of step budgets that spanned from~$10^3$ to~$10^8$ and selected the results for the largest budget for which we obtained results in a timely fashion.

To evaluate our approach, we run our experiments on SUTs simulated from Mealy machines (stored as DOT files) obtained with automata learning in previous work (see \autoref{sec:case-studies}). This way our experiments are not hindered by e.g. response time of a real SUT. For each case study, multiple SUT models were available, as they were obtained from different real implementations. All properties available for the case study were converted from their format, which ranged from  natural language to LTL formulae, to DFAs. Additionally, we also created a global property for the case study by applying conjunction on all properties available for the case study. Experiments were performed for all combinations of SUTs and properties.

We took the following decisions in obtaining the case studies:
\begin{description}
 \item[BLE] We used all available SUTs and properties.
 \item[SSH (LTL)] We use nine of the LTL properties for our evaluation. Property 3, property 4 and property 9 are excluded, because they each use either NuSMV's~$O$ (once), or NuSMV's~$S$ (since) LTL operator, neither of which are supported by Spot\footnote{\url{https://spot.lre.epita.fr/concepts.html\#ltl}}.
 \item[SSH (DFA) \& DTLS (Client \& Server)]
 We restricted ourselves to the SUTs for which Fiterau et al. included bug reports, i.e. those covered in the paper.
 \item[RERS] We used the thirty provided models, and the safety properties among their accompanying sets of twenty LTL properties. The number of safety properties, and thereby of the properties that we used per model ranges from 9 to 18, with a mean of 13.7 and a standard deviation of 1.9.
\end{description}
We performed the black-box checking experiments on a computer with an AMD Epyc 7642 processor. Every experiment was constrained to 2 processing cores and 4Gb of RAM.

\subsection{Model-based Testing}
In order to evaluate BBC's performance against a baseline, we applied (standalone) Model-Based Testing on both the BLE, and SSH (LTL) benchmark implementations, as well as a subset of the RERS benchmark. We chose these case studies as they are among the simplest, and their specifications are small enough to be handled by the tool in a timely fashion.

For the MBT implementation, we use Lattest \footnote{\url{https://github.com/ramon-janssen/lattest}}, a Haskell library for MBT. As a test selection strategy, we implement a tester with memory: across the test suite, the input sequences that were tested already for each state are stored, and, at each step, a new input is chosen randomly from the set of inputs that has not been tried yet. This way we ensure variation of inputs to increase effectiveness for detecting bugs.

Test execution is performed separately for each property. Earlier attempts to test the conjunction of all properties performed poorly, as shallow bugs were reported too often, making it more difficult to detect deeper ones.

For each case study, the number of test cases for the test suites is ten times the number of queries it takes for the BBC implementation to find the corresponding bug. The number of test steps, within those tests, is defined as twice the number of states of the specification, to aid the identification of deeper bugs. Each test suite is repeated fifty times, with different seeds.

The specifications for these experiments were constructed from LTL formulae as described in \autoref{def specification}. As in BBC, the implementations described in the DOT files are simulated by a Python script. Both the tester and the Python-based SUT are executed inside Docker containers.

MBT experiments were executed on a computer with an Intel i7 processor. Each instance of the tester was constrained to 1 processing core and 2Gb of RAM.


\section{Results}
In this section, we present the results of our experiments, addressing each of the questions posed in \autoref{sc:intro}.

\subsection{BBC effectiveness evaluation}
\begin{figure}[t!]
    \includegraphics[width=1.1\linewidth]{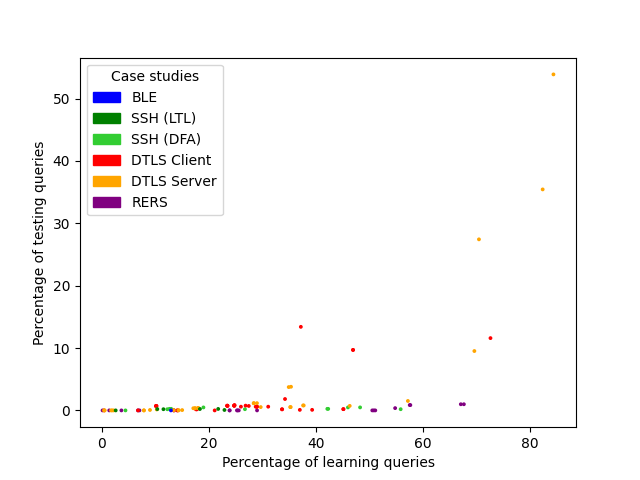}
    \caption{Scatter plot showing the percentages of the number of learning and testing queries needed for BBC to find a bug, relative to the total number of learning and testing queries needed to learn the full model.}
    \label{fig:scatter_bbc_vs_learning}
\end{figure}
First, we evaluate the effectiveness of model checking all hypotheses in BBC. \autoref{fig:scatter_bbc_vs_learning} compares the efficiency of BBC with that of active model learning. Each data point represents a benchmark, a combination of an SUT and a property for which we found a counterexample with BBC. On the x-axis, we have the percentage of the number of learning queries that we needed to find a counterexample with BBC, relative to the number of learning queries needed to find a counterexample with model learning. The y-axis shows this percentage for the required number of testing queries. We needed an average of 27.2\% of the number of learning queries and 1.9\% of the number of testing queries for BBC relative to model learning. When taking all queries together, we needed an average of 3.4\% of the queries for BBC as we needed for model learning.

\autoref{tab:counterexes_table_no-step-budgets} indicates how successful we were at finding counterexamples in the experiments in which we didn't use step budgets. For every SUT, we have the number~$t$ of specifications that we have in total, the number~$c \leq t$ of specifications that aren't satisfied by the SUT, and the number~$f \leq c$ of specifications for which we found a counterexample for this SUT. For all SUTs in~\autoref{tab:counterexes_table_no-step-budgets}, $f = c$. We indicate the values of~$f$ and~$t$ for these SUTs in the form~$f/t$ in the final column. \autoref{tab:counterexes_table_only-step-budgets} likewise provides the values of~$f$, $c$ and~$t$ for the SUTs for which we did use step budgets, in the form~$f/c/t$. The table indicates that we didn't find all counterexamples for the SUTs for which we used step budgets. This is a consequence of using such budgets, since learning a model incompletely incurs the risk that not all bugs are found. Still, we can read from \autoref{tab:counterexes_table_only-step-budgets} that for instance for the RERS case study, we found $94\%$ of the safety bugs across all SUTs. \autoref{tab:counterexes_table_no-step-budgets} and \autoref{tab:counterexes_table_only-step-budgets} collectively indicate the sizes of the state and input sets of all SUTs that we covered in our experiments. \autoref{tab:counterexes_table_only-step-budgets} additionally specifies the average sizes of the state sets of the final hypotheses, the average number of queries that we required for the experiments when we used BBC with runtime monitoring, and the percentage of the queries that we needed when we use runtime monitoring for BBC, compared to when we didn't.

\subsection{BBC vs. MBT}
\begin{figure}[ht]
    \includegraphics[width=1.1\linewidth]{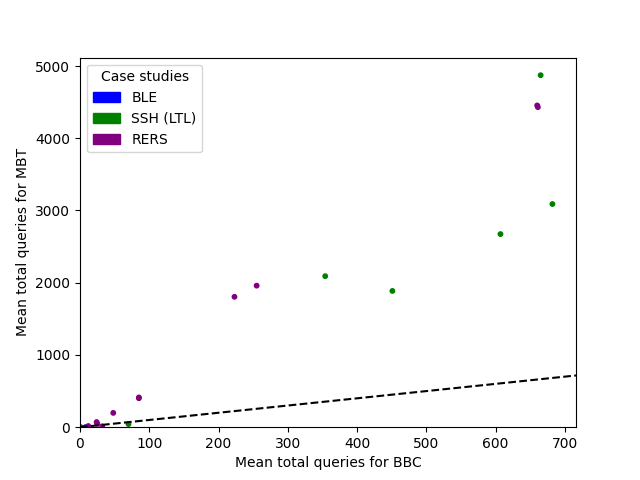}
    \caption{Scatter plot showing the mean number of queries needed for BBC to find a bug, compared to the mean number of tests MBT needs to find the same bug.}
    \label{fig:scatter_bbc_vs_mbt}
\end{figure}

Finally, we compare BBC's performance for bug finding with standalone MBT, for a subset of the benchmarks.

 \autoref{tab:findings} compares how frequently bugs were reported using BBC and MBT. With BBC, all violations to properties were reported in all experiments repetitions. In contrast, MBT reported bugs for approximately 70\% of the evaluated properties, out of which roughly 26\% were reported consistently across all experiments.

\autoref{fig:scatter_bbc_vs_mbt} compares the number of tests needed to find the counterexamples with MBT with the total number of queries required to find counterexamples for these benchmarks with BBC. We needed an average of 260 queries to find the counterexamples for BBC and an average of 1498 tests to find the counterexamples with MBT.

\section{Conclusions and Future Work}

With a systematic evaluation, we confirmed the intuition of Peled, Vardi \& Yannakakis \cite{PeledVY99,PeledVY02} that bugs are detected (much) faster when, as part of the BBC loop, model checking is applied for all hypotheses, rather than just for the final hypothesis.
We expect that this conclusion is robust and will also apply for other benchmarks and for different algorithms for model learning and model-based testing, although this needs to be confirmed by further studies.
We also found that adding runtime monitoring to the BBC toolbox leads to a small decrease (5\%) of the number of queries required to detect bugs.  In our experiments we only applied runtime monitoring for the learning queries. It is future work to explore whether application to testing queries will lead to further speedups.

When the full model cannot be learned, BBC is still able to detect many violations of the specification. In particular, using BBC, we managed to detect 94\% of the safety properties violations in the challenging RERS 2019 industrial LTL benchmarks.
This improves (what we believe to be) the best previous result for this benchmark collection by Kruger, Junges \& Rot \cite{KrugerJR24}, who succeeded to learn partial models for 23 benchmarks (they did not consider the 7 largest benchmarks), and full models for 15  benchmarks.

Our results show that, when the objective of model learning is to detect bugs in implementations, it really helps to have a formal specification available that may serve as input for a BBC loop.
Even in cases where model learning is unable to learn the full SUT model, BBC is able to learn a small hypothesis model that is still large enough to allow for the detection of violations of the specification.
In one of our benchmarks (RERS M65), for example, the full SUT model has 3966 states, but hypotheses models with (on average) 29 states are enough to detect that the SUT violates 5 out of the 17 properties in the specification.
This provides a nice illustration of the aphorism attributed to statistician George Box that ``all models are wrong but some are useful.''

Our results also confirm the frequent observation that BBC is way more effective than existing MBT algorithms in finding deep bugs in implementations.
As observed by Vaandrager \& Melse \cite{VM25}, prominent MBT approaches such as
the W-method of Vasilevskii \cite{Vas73} and Chow \cite{Ch78},
the Wp-method of Fujiwara et al \cite{FujiwaraEtAl91}, and
the HSI-method
of Luo et al \cite{Luo1995} and Petrenko et al \cite{YP90,PetrenkoYLD93}, are complete for 
fault domains that contain all FSMs in which any state can be reached by first performing a sequence from a state cover for the specification model, followed by up to $k$ arbitrary inputs, for some small $k$.
This means that when all states of the SUT model can be reached by first doing an access sequence for some state of the specification, followed by a few arbitrary inputs, then these MBT approaches are guaranteed to find bugs in this SUT (when present). Outside these fault domains they are much less effective. Our benchmarks have up to 200 inputs, which means that we can only run the MBT algorithms for values of $k$ up to $2$.  We conjecture that for many of our benchmarks it is not possible to reach all SUT states by an access sequence of a specification DFA state followed by at most $2$ arbitrary inputs.  Exploring this conjecture (which might explain the limitations of existing MBT algorithms) is a question for future research.


%
%
%
%
\bibliographystyle{splncs04}
\bibliography{abbreviations,references,bibliography}

%
%
\appendix

\newpage

\section{Appendix}

\begin{table}[H]
	\begin{tabular*}{\linewidth}{@{\extracolsep{\fill}} lrrr }
\toprule
SUT & $\vert Q \vert$ & $\vert I \vert$ & \# Props \\
\midrule
\textbf{BLE} &  &  &  \\
cc2640r2f-no-feature & 11 & 8 & 0/1 \\
cc2640r2f-no-length & 11 & 8 & 0/1 \\
cc2640r2f-no-pairing & 6 & 8 & 0/1 \\
cc2650 & 5 & 9 & 0/1 \\
cc2652r1 & 4 & 7 & 1/1 \\
cyble-416045-02 & 3 & 9 & 0/1 \\
cyw43455 & 16 & 7 & 0/1 \\
nRF52832 & 5 & 9 & 0/1 \\
\midrule
\textbf{DTLS Client} &  &  &  \\
etinydtls\_client\_ecdhe\_cert & 17 & 17 & 2/16 \\
gnutls-3.6.7\_client\_dhe\_ecdhe\_rsa\_cert\_reneg & 27 & 21 & 3/16 \\
jsse-12.0.2\_client\_ecdhe\_cert & 133 & 17 & 7/16 \\
jsse-16.0.1\_client\_ecdhe\_cert & 113 & 17 & 7/16 \\
mbedtls-2.16.1\_client\_dhe\_ecdhe\_rsa\_cert\_reneg & 34 & 21 & 2/16 \\
piondtls-2.0.9\_client\_ecdhe\_cert & 26 & 16 & 1/16 \\
piondtls-usenix\_client\_ecdhe\_cert & 49 & 16 & 10/16 \\
scandium-2.0.0-M16\_client\_ecdhe\_cert & 38 & 17 & 7/16 \\
scandium-2.6.2\_client\_ecdhe\_cert & 12 & 17 & 0/16 \\
wolfssl-4.0.0\_client\_psk & 13 & 8 & 1/16 \\
wolfssl-4.7.1r\_client\_psk & 17 & 8 & 1/16 \\
\midrule
\textbf{DTLS Server} &  &  &  \\
ctinydtls\_ecdhe\_cert\_req & 30 & 10 & 2/18 \\
etinydtls\_ecdhe\_cert\_req & 27 & 10 & 2/18 \\
gnutls-3.5.19\_all\_cert\_req & 43 & 16 & 2/18 \\
gnutls-3.6.7\_all\_cert\_req & 16 & 16 & 1/18 \\
gnutls-3.7.1\_all\_cert\_req & 16 & 16 & 1/18 \\
jsse-12.0.2\_ecdhe\_cert\_req & 124 & 10 & 9/18 \\
jsse-16.0.1\_ecdhe\_cert\_req & 78 & 10 & 0/18 \\
mbedtls-2.16.1\_all\_cert\_req & 17 & 16 & 1/18 \\
mbedtls-2.26.0\_all\_cert\_req & 17 & 16 & 1/18 \\
openssl-1.1.1b\_all\_cert\_req & 19 & 16 & 5/18 \\
openssl-1.1.1k\_all\_cert\_req & 23 & 16 & 5/18 \\
piondtls-2.0.9\_ecdhe\_cert\_req & 25 & 10 & 0/18 \\
piondtls-usenix\_ecdhe\_cert\_req & 66 & 10 & 4/18 \\
scandium-2.6.2\_ecdhe\_cert\_req & 11 & 10 & 1/18 \\
wolfssl-4.0.0\_psk & 10 & 7 & 0/18 \\
wolfssl-4.7.1r\_psk & 13 & 7 & 0/18 \\
\midrule
\textbf{SSH (DFA)} &  &  &  \\
BitVise-8.49 & 43 & 12 & 4/16 \\
DropBear-v2020.81 & 21 & 12 & 2/16 \\
OpenSSH-8.8p1 & 37 & 12 & 5/16 \\
\midrule
\textbf{SSH (LTL)} &  &  &  \\
BitVise-7.23 & 66 & 13 & 2/9 \\
DropBear-v2014.65 & 29 & 13 & 1/9 \\
OpenSSH-6.9p1-2 & 31 & 21 & 3/9 \\
\bottomrule
\end{tabular*}

	\caption{The number of properties found with BBC, compared to the total number of properties for each SUT, without the use of step budgets}
	\label{tab:counterexes_table_no-step-budgets}
\end{table}%

\begin{table}[H]
	\begin{tabular*}{\linewidth}{@{\extracolsep{\fill}} lrrrrrrlr }
\toprule
SUT & $\vert Q \vert$ & $\vert I \vert$ & \multicolumn{ 2 }{l}{ $\vert Q^\mathcal{H} \vert$ } & \multicolumn{ 2 }{l}{ \# Qs req. } & Qs needed of n- & \# Props \\
\cmidrule(lr){4-5}\cmidrule(lr){6-7}
&&&avg & sd&avg & sd&on. ob tree mon.& \\
\midrule
\textbf{DTLS Client} (s. bud.: 1e6) &  &  &  &  &  &  \\
Openssl-1.1.1B\_Client ... & 387 & 21 & 17 & 3 & 24408 & 22018 & 100.0\% & 9/9/16 \\
Openssl-1.1.1K\_Client ... & 387 & 21 & 17 & 3 & 24408 & 22018 & 100.0\% & 9/9/16 \\
\midrule
\textbf{DTLS Client} (s. bud.: 1e8) &  &  &  &  &  &  \\
Ctinydtls\_Client\_Ecdh ... & 22 & 17 & 14 & 0 & 428 & 457 & 100.0\% & 2/2/16 \\
Gnutls-3.7.1\_Client\_D ... & 39 & 21 & 6 & 0 & 446 & 512 & 100.0\% & 3/3/16 \\
Mbedtls-2.26.0\_Client ... & 40 & 21 & 34 & 0 & 387 & 235 & 100.0\% & 2/2/16 \\
\midrule
\textbf{DTLS Server} (s. bud.: 5e4) &  &  &  &  &  &  \\
Scandium-2.0.0-M16\_Ec ... & 1007 & 10 & 46 & 34 & 3949 & 826 & 100.0\% & 9/10/18 \\
\midrule
\textbf{RERS} (s. bud.: 1e5) &  &  &  &  &  &  \\
M65 & 3966 & 33 & 29 & 10 & 3788 & 4738 & 100.0\% & 5/5/17 \\
\midrule
\textbf{RERS} (s. bud.: 1e7) &  &  &  &  &  &  \\
M131 & 1017 & 181 & 60 & 16 & 73700 & 103969 & 95.8\% & 2/3/13 \\
M132 & 2441 & 190 & 1 & 0 & 97 & 136 & 25.3\% & 2/3/13 \\
M181 & 1347 & 93 & 92 & 36 & 279169 & 275537 & 99.9\% & 6/6/16 \\
M182 & 657 & 75 & 68 & 40 & 112640 & 153017 & 99.8\% & 2/2/15 \\
M190 & 456 & 52 & 92 & 50 & 116571 & 91762 & 104.6\% & 8/8/13 \\
M22 & 93 & 108 & 55 & 14 & 193110 & 334382 & 100.0\% & 3/8/12 \\
M24 & 284 & 100 & 0 & 0 & 128086 & 220164 & 94.2\% & 7/7/15 \\
\midrule
\textbf{RERS} (s. bud.: 1e8) &  &  &  &  &  &  \\
M106 & 79 & 83 & 48 & 8 & 170699 & 272543 & 99.9\% & 7/7/13 \\
M135 & 57 & 177 & 0 & 0 & 423994 & 506331 & 91.1\% & 6/6/13 \\
M158 & 28 & 120 & 6 & 1 & 206223 & 151881 & 98.8\% & 5/5/12 \\
M159 & 30 & 120 & 21 & 4 & 198862 & 196726 & 98.1\% & 7/7/12 \\
M164 & 43 & 102 & 31 & 12 & 141351 & 103192 & 91.1\% & 11/11/18 \\
M167 & 163 & 152 & 98 & 41 & 362042 & 389086 & 95.5\% & 5/5/13 \\
M172 & 113 & 98 & 19 & 7 & 119188 & 78982 & 93.0\% & 5/5/13 \\
M173 & 483 & 27 & 23 & 6 & 6688 & 10174 & 95.9\% & 5/6/14 \\
M183 & 9 & 12 & 9 & 0 & 101 & 70 & 74.4\% & 8/8/16 \\
M185 & 190 & 71 & 31 & 3 & 63138 & 64292 & 93.5\% & 4/4/9 \\
M189 & 289 & 138 & 61 & 16 & 1189372 & 1357323 & 99.9\% & 6/7/16 \\
M196 & 81 & 24 & 17 & 2 & 3596 & 5732 & 99.8\% & 5/5/13 \\
M199 & 27 & 56 & 1 & 0 & 42408 & 94629 & 99.1\% & 8/8/13 \\
M201 & 128 & 87 & 66 & 9 & 506915 & 877413 & 99.9\% & 3/3/14 \\
M27 & 198 & 201 & 34 & 20 & 1649449 & 731451 & 99.7\% & 7/8/15 \\
M41 & 25 & 47 & 25 & 1 & 10069 & 6474 & 96.2\% & 5/5/11 \\
M45 & 184 & 32 & 63 & 4 & 24196 & 41878 & 76.5\% & 3/3/14 \\
M49 & 142 & 75 & 106 & 14 & 89322 & 135360 & 103.9\% & 4/4/13 \\
M54 & 27 & 10 & 15 & 2 & 365 & 298 & 96.1\% & 12/12/16 \\
M55 & 181 & 156 & 31 & 10 & 299926 & 401222 & 100.0\% & 10/10/13 \\
M76 & 210 & 26 & 5 & 4 & 22972 & 45027 & 99.1\% & 4/5/14 \\
M95 & 33 & 99 & 24 & 7 & 3785 & 3306 & 99.5\% & 6/6/11 \\
\bottomrule
\end{tabular*}

	\caption{The number of properties found with BBC, compared to the total number of properties for each SUT, where (varying) step budgets are used}
	\label{tab:counterexes_table_only-step-budgets}
\end{table}%

\begin{table}[t]
	\centering
	\begin{tabular*}{\linewidth}{@{\extracolsep{\fill}} llcc }
	\toprule
	System & Property & MBT & BBC \\
	\midrule
	
	BLE/cc2652r1f & Prop 0  & 50/50 & 50/50 \\
	
	\midrule
	
	\multirow{3}{*}{SSH/OpenSSH}
	& Prop 5  & 22/50 & 50/50 \\
	& Prop 8  & 15/50 & 50/50 \\
	& Prop 11 & 26/50 & 50/50 \\
	
	\midrule
	
	\multirow{2}{*}{SSH/BitVise}
	& Prop 8  & 50/50 & 50/50 \\
	& Prop 11 & 5/50  & 50/50 \\
	
	\midrule
	
	SSH/DropBear
	& Prop 12 & 6/50  & 50/50 \\
	
	\midrule
	
	\multirow{8}{*}{RERS/m183}
	& Prop 1  & 0/50  & 50/50 \\
	& Prop 2  & 0/50  & 50/50 \\
	& Prop 7  & 17/50 & 50/50 \\
	& Prop 9  & 23/50 & 50/50 \\
	& Prop 12 & 50/50 & 50/50 \\
	& Prop 15 & 0/50  & 50/50 \\
	& Prop 16 & 23/50 & 50/50 \\
	& Prop 17 & 50/50 & 50/50 \\
	
	\midrule
	
	\multirow{12}{*}{RERS/m54}
	& Prop 0  & 2/50  & 50/50 \\
	& Prop 3  & 0/50  & 50/50 \\
	& Prop 4  & 50/50 & 50/50 \\
	& Prop 5  & 2/50  & 50/50 \\
	& Prop 6  & 50/50 & 50/50 \\
	& Prop 7  & 0/50  & 50/50 \\
	& Prop 8  & 50/50 & 50/50 \\
	& Prop 9  & 1/50  & 50/50 \\
	& Prop 11 & 0/50  & 50/50 \\
	& Prop 13 & 0/50  & 50/50 \\
	& Prop 15 & 2/50  & 50/50 \\
	& Prop 18 & 0/50  & 50/50 \\
	\bottomrule
\end{tabular*}
	\caption{Comparison of bug reports for MBT and BBC across systems and properties.}
	\label{tab:findings}
\end{table}

\end{document}